\documentstyle[11pt]{article}
\newcommand{\beq}{\begin{equation}}
\newcommand{\eeq}{\end{equation}}
\newcommand{\dd}[2]{\frac {\partial #1}{\partial #2}}

\newcommand{\beqs}{\begin{eqnarray}}
\newcommand{\eeqs}{\end{eqnarray}}

\newcommand{\nn}{\nonumber}

\begin{document}

\begin{center}
\vskip 2.5cm
{\LARGE \bf Path integral treatment of two- and three- dimensional
delta-function potentials and application to spin-1/2 Aharonov-
Bohm problem}
\vskip 1.0cm
{\Large  D.~K.~Park* }
\\
{\large  Department of Physics and Astronomy, University of Rochester\\
Rochester, NY 14627 USA}

\vskip 0.4cm
\end{center}

\centerline{\bf Abstract}

Delta-function potentials in two- and
three-dimensional quantum mechanisc are
analyzed by the incorporation of the self-adjoint
extension method to the path integral formalism.
The energy-dependent Green functions for free particle
plus delta-function potential systems are explicitly calculated.
Also the energy-dependent Green function for the
spin-1/2 Aharonov-Bohm problem is evaluated. It is found
that the only one special value of the self-adjoint
extension parameter gives a well-defined and non-trivial
time-dependent propagator. This special value corresponds
to the viewpoint of the spin-1/2 Aharonov-Bohm problem
when the delta-function is treated as a limit of the
infinitesimal radius.

\vfill
\begin{center}
March 1994
\end{center}

-------------

* On leave from Department of Physics,KyungNam University, Masan,
631-701, Korea
\newpage
\setcounter{footnote}{1}

\newcommand{\tr}{\;{\rm tr}\;}
\section{Introduction and Basic tool for the calculation}

Recently two-dimensional delta-function potential has been of
interest in the problems related to the Aharonov-Bohm(AB)
effect of spin-1/2 particles \cite{A, B} in which the delta-
function is a mathematical description of Zeeman interaction
of the spin with magnetic flux tube. The higher dimensional
delta-function potential also appears in the
non-relativistic limit of $\phi^{4}$ scalar field theory.\cite{C}
In this case there arise ultraviolet divergences even though
one is dealing with quantum mechanics and renomalization is
required. B\'{e}g and Furlong concluded that the delta-function
interaction gives rise to a trivial S-matrix when the coupling
constant is finite. Recently a proposal was made for defining
a non-trivial delta-function interaction in two and three
dimensions.\cite{D} It was found in Ref.[4] that in order
for the delta function interaction to give a non-trivial effect
the bare coupling constant should be infinite and all the physical
quantities should be expressed as a renomalized coupling
constant which is regarded as finite quantity from the usual viewpoint
of a renomalization theory. The infinite quantity of the bare
coupling constant of two-dimensional delta-functional potential
is realized in the spin-$\frac{1}{2}$ AB problem.

Jackiw also presented another pure mathmatical method, self-
adjoint extension method \cite{E, F}, which also makes, in addition
to the renomalization method, the delta-function interaction to be
finite and non-trivial. In the self-adjoint extension
method the delta-function potential is categorized as a
solvable model.\cite{F} Under the assumption of the finiteness
of the self-adjoint extension parameters which naturally arise
in the self-adjoint extension method since a n-parameter family of
solutions is yielded by this method, one can make the physical
quantities to be finite and non-trivial. Also it is proved
that even though the latter is pure mathematically-based
method, it gives a result identical to that of the
renomalization method if a certain relation between the
self-adjoint extension parameter and renomalized(or bare)
coupling constant is required. This relation is derived by
identifying the scattering amplitudes and bound state energies
which are calculated by these two different methods respectively.
It will be shown in this paper that the relation of the self-adjoint
extension parameter with the coupling constant is directly
and more consistently derived in the course of formulation
by path integral method.

Recently the path integration of the one dimensional delta-
function potential, in this case self-adjoint extension is
not required, is exactly calculated \cite{G, H, I} by making
use of the Brownian amplitude. More recently by the
incorporation of Dirichlet and Neumann boundary conditions
into the path integral formalism the closed expression for
the energy-dependent Green functions are obtained \cite{J, K}
via summation of perturbation expansions when the potentials
are one-dimensional $\delta(x)$ and $\delta^\prime(x)$
respectively.
In this paper the energy-dependent Green functions will be
explicitly calculated when the potentials are two- and three-
dimensional delta-functions by the incorporation of self-adjoint
extension method into the path integral formalism. For this
purpose the technique used by Gaveau and Schulman in Ref.[7]
is developed as follows.

Consider a d-dimensional system whose Hamiltonian is
\beq
     H = H_{V}({\bf p}, {\bf r}) + v \delta({\bf r})
\eeq
where
\beq
     H_{V} = \frac{{\bf p}^2}{2M} + V({\bf r}).
\eeq
Then the time-dependent Brownian propagator $G[{\bf r_{1}}, {\bf r_{2}}
;t]$ for the Hamiltonian H obeys the integral equation \cite{L, M}
\beq
     G[{\bf r_{1}}, {\bf r_{2}};t] = G_{V}[{\bf r_{1}}, {\bf r_{2}}
;t] - v \int_{0}^{t} ds \int d{\bf r} G_{V}[{\bf r_{1}}, {\bf r}
;t-s] \delta({\bf r}) G[{\bf r}, {\bf r_{2}};s]
\eeq
where $G_{V}[{\bf r_{1}}, {\bf r_{2}};t]$ is time-dependent Browian
propagator for the Hamiltonian $H_{V}$. A Feynman propagator(or
Kernel) $K[{\bf r_{1}}, {\bf r_{2}};t]$ is straightforwardly evaluated
from a Brownian propagator by the relation
\beq
    K[{\bf r_{1}}, {\bf r_{2}};t] = G[{\bf r_{1}}, {\bf r_{2}};it].
\eeq
By taking a Laplace transform
\beq
    \hat{f} (E) \equiv L{f(t)} \equiv \int_{0}^{\infty} dt
     e^{-Et} f(t)
\eeq
to the both sides of Eq.(3), the relation
\beq
    \hat{G} [{\bf 0}, {\bf r_{2}};E] = \frac{\hat{G}_{V} [{\bf 0},
    {\bf r_{2}};E]}{1 + v \hat{G}_{V} [{\bf 0}, {\bf 0};E]}
\eeq
is obtained. By combining Eqs.(3) and (6) the energy-dependent
Brownian Green function  $\hat{G} [{\bf r_{1}}, {\bf r_{2}};E]$
is easily calculated. The usual energy-dependent Green function
$\hat{K} [{\bf r_{1}}, {\bf r_{2}}, E]$ which is a Fourier transform
of $K[{\bf r_{1}}, {\bf r_{2}};t] \theta(t)$, where $\theta(t)$ is a
step function,  is also straightforwardly evaluated from $\hat{G}[{
\bf r_1}, {\bf r_2};E]$ by the relation
\beq
     \hat{K} [{\bf r_{1}}, {\bf r_{2}};E] = -i
     \hat{G} [{\bf r_{1}}, {\bf r_{2}};-E].
\eeq
Also by taking a inverse Laplace transform to $\hat{G}[{\bf r_1},
{\bf r_2};E]$ one can evaluate the time dependent Brownian
propagator $G[{\bf r_{1}}, {\bf r_{2}};t]$ and Feynman propagator
$K[{\bf r_{1}}, {\bf r_{2}};t]$.
This is one of the methods used by Gaveau and Schulman in
Ref.[7] for the calculation of energy-dependent Green function
and time-dependent propagator when the potential is one-
dimensional delta function. When $V(x) = 0$, the energy-dependent
Green function and the time-dependent propagator are respectively
\beq
     \hat{G} [x, y, E] = \frac{e^{-\sqrt{2E}\mid x-y \mid}}
     {\sqrt{2E}} - v \frac{e^{-\sqrt{2E}(\mid x \mid +
     \mid y \mid)}}{\sqrt{2E} (\sqrt{2E} + v)}
\eeq
and
\beq
     G[x, y; t] = G_{0}[x, y;t] - v\int_{0}^{\infty}dz
     e^{-vz} G_{0}[\mid x \mid, -\mid y \mid - \mid z
     \mid;t],
\eeq
where $G_0[x, y;t]$ is time-dependent propagator for the
one-dimensional free particle.
Since $G[x, y;t] = \sum_{E} e^{-Et} \phi_{E}(x) \phi_{E}^{*}(y)$
 , where $\phi_{E}(x)$ is a eigenfunction of a Hamiltonian
operator, $G[x, y;t]$ must obey the same boundary condition with
that of the eigenfunction at the origin which arises because of
the singular delta function potential. It is easily proved from
Eq.(9) that $G[x, y;t]$ obeys the boundary condition
\beq
    \dd{G}{x}(0^{+}, y;t) - \dd{G}{x}(0^{-}, y;t)
    = 2 v G[0, y;t].
\eeq
The important fact that will play the crucial role for the
calculation of higher-dimensional delta-function potential
case is that there are two different ways for the calculation of
the bound state energy from Eq.(8). First the bound state
energy can be obtained by the pole of $\hat{G} [x, y;E]$. This is
a univertial property of a energy-dependent Green function.
Second the bound state energy can be calculated by applying
the boundary condition(10) to $\hat{G} [x, y;E]$
\beq
     \dd{ \hat{G} }{x}[0^{+}, y;E] - \dd{ \hat{G} }{x}[0^{-}, y;
     E] = 2 v \hat{G} [0, y;E].
\eeq
This means that the boundary condition(10) plays an
important role for the occurrence of the bound state.
One can prove that the bound state energies which are
calculated from both cases respectively are exactly same.

The above-mentioned method, however, cannot be applied
directly to the higher dimensional delta-function potential
system because the energy-dependent Green function of free
particle system  $\hat{G}_{0}[{\bf x}, {\bf y};E]$ diverges
when ${\bf x}$ is contiguous to ${\bf y}$.
So in this case in order to escape the
infinity Eq.(6) will be modified by
\beq
     \hat{G} [{\bf 0}, {\bf r_{2}};E] = \frac{
     \hat{G}_{V} [{\bf 0}, {\bf r_{2}};E]}{1 +
     v \lim_{\epsilon \rightarrow 0^{+}}
     \hat{G}_V [{\bf \epsilon}, {\bf 0};E]}.
\eeq
By combining Eqs.(3) and (12)  the energy-dependent
Green function for this case is
\beq
     \hat{G} [{\bf r_{1}}, {\bf r_{2}};E] -
     \hat{G}_{V} [{\bf r_{1}}, {\bf r_{2}};E] = -
     \frac{\hat{G}_{V} [{\bf r_{1}}, {\bf 0};E] \hat{G}_{V} [
     {\bf 0}, {\bf r_{2}};E]}{ \frac{1}{v} + \lim_{
    \epsilon \rightarrow 0^{+}} \hat{G}_{V} [{\bf \epsilon},
     {\bf 0};E]}.
\eeq
By applying the boundary condition at the origin which
arises by the self-adjoint extension of $H_{V}$ to $
\hat{G} [{\bf r_{1}}, {\bf r_{2}};E]$ one can calculate
directly the bound state energy. From the fact that this
bound state energy must be a pole of the right-hand side
of Eq.(13) the relation between the self-adjoint
extension parameter and bare coupling constant can be
derived. By using this relation $\hat{G} [{\bf r_{1}},
 {\bf r_{2}};E] $ can be expressed in terms of the
self-adjoint extension parameters.
The more serious situation arises when the above-mentioned
method is applied to the spin-1/2 AB problem. Since, in
this case, $H_V$ is Hamiltonian of usual spin-0 AB problem,
both $\hat{G}_V[{\bf 0}, {\bf r};E]$ and
$\hat{G}_V[{\bf r}, {\bf 0};E]$ are not well-defined because
of the magnetic flux tube at the origin.
So Eq.(13) must
be modified again by
\beq
     \hat{G} [{\bf r_{1}}, {\bf r_{2}}, E]
     - \hat{G}_{V} [{\bf r_{1}}, {\bf r_{2}};E]
     = - \frac{\hat{G}_{V} [{\bf r_{1}}, {\bf
     \epsilon_{1}};E] \hat{G}_{V} [{\bf \epsilon_{1}},
     {\bf r_{2}};E]}{\frac{1}{v} + \lim_{\epsilon
     _{2} \rightarrow \epsilon_{1}^{+}} \hat{G}
     _{V} [{\bf \epsilon_{2}}, {\bf \epsilon_{1}};E]}.
\eeq
The remaining calculation technique is exactly same
with the previous case.

It is worthwhile to note that the relation between
the self-adjoint extension parameter and the bare
coupling constant is naturally derived in the frame
of the path integral formulation. In the usual
quantum mechanics this relation is derived only by
identifying the various physical quantities which
are calculated by renomalization and self-adjoint
extension methods respectively.
This means this relation is derived in path integral
formulation more consistently and convincingly.

This paper is organized as follows. In Sec.2 the
path integral formulation for the two- and three-
dimensional delta-function potentials are analyzed
when $V({\bf r}) = 0$. The energy-dependent Green
functions are explicitly calculated. It is found
that the time-dependent propagator of two-dimensional
case is not well-defined while that of three-
dimensional case is well-defined. In Sec.3 two-dimensional
spin-1/2 AB problem is formulated
by the path integral method. It is found that the only
one special value of self-adjoint extension parameter
gives a well-defined and non-trivial time-dependent
propagator. This special value coincides with the
result of Ref.[1] in which the delta-function is
treated as a limit of the infinitesimal radius developed
by Hagen. Although it is known that
there is only one Hamiltonian that is correct limit
as the solenoid radius goes to zero\cite{N}, the conviction
is far from universial. I think the fact of the existence
of the time dependent propagator only in that case casts
the another clue to prove the universality. In Sec.4 the
brief conclusion is given.

\section{Path integral formulation of two- and three-
dimensional delta-function potentials}

In this section the energy-dependent Green functions
for the two- and three-dimensional delta-function
potentials will be calculated explicitly.

Consider the Hamiltonian
\beq
     H_{d} = \frac{{\bf p^{2}}}{2M} + v \delta({\bf r})
\eeq
where d is a dimension of space. Since the propagator
of the free particle in d dimension is well-known as
\beq
     G_{0}^{(d)}[{\bf r_{1}}, {\bf r_{2}};t] =
     \left(\frac{M}{2\pi t}\right)^{\frac{d}{2}}
     e^{-\frac{M}{2t} ({\bf r_{1}} - {\bf r_{2}})^{2}},
\eeq
one can calculate the energy-dependent Green function for
the free particle system by using a  Laplace
transformation
\beqs
      \hat{G}_{0}^{(2)} [{\bf r_{1}}, {\bf r_{2}};E]
      = \frac{M}{\pi} K_{0}[\sqrt{2ME}\mid {\bf r_{1}}
      - {\bf r_{2}} \mid], \nn \\
      \hat{G}_{0}^{(3)} [{\bf r_{1}}, {\bf r_{2}};E]
      = \frac{M}{2\pi} \frac{e^{-\sqrt{2ME}\mid
      {\bf r_{1}} - {\bf r_{2}}\mid}}
      {\mid{\bf r_{1}} - {\bf r_{2}}\mid}
\eeqs
where $K_{\nu}$ is usual modified Bessel function.
Note that  $\hat{G}_{0}^{(d)} [{\bf 0}, {\bf 0};E]$
goes to the infinity as stated in the previous section.
By inserting Eq.(17) into Eq.(13) one gets
\beqs
     \hat{A}^{(2)} [{\bf r_{1}}, {\bf r_{2}};E]
       = -\frac{1}{\frac{1}{v} + \frac{M}{\pi}
       K_{0}[\sqrt{2ME} \epsilon]} \hat{G}_{0}^{(2)}
       [{\bf r_{1}}, {\bf 0};E] \hat{G}_{0}^{(2)} [
       {\bf 0}, {\bf r_{2}};E],   \nn \\
       \hat{A}^{(3)} [{\bf r_{1}}, {\bf r_{2}};E]
       = -\frac{1}{\frac{1}{v} + \frac{M}{2 \pi
       \epsilon} e^{-\sqrt{2ME}\epsilon}}
       \hat{G}_{0}^{(3)} [{\bf r_{1}}, {\bf 0};E]
       \hat{G}_{0}^{(3)} [{\bf 0}, {\bf r_{2}};E]
\eeqs
where
\beq
       \hat{A}^{(d)} [ {\bf r_{1}}, {\bf r_{2}};E] =
       \hat{G}^{(d)} [{\bf r_{1}},
       {\bf r_{2}};E] - \hat{G}_{0}^{(d)} [{\bf
       r_{1}}, {\bf r_{2}};E].
\eeq
By applying the boundary condition at the origin
which arises by the self-adjoint extension of
Hamiltonian
\beqs
\lefteqn{ \lim_{r_{1} \rightarrow 0}
        \frac{\hat{A}^{(2)} [{\bf r_{1}}, {\bf r_{2}}; E]}
             { \ln r_{1} } } \nn \\
  & &         =   \frac{\lambda}{\pi}
                  \lim_{r_{1} \rightarrow 0}
                 \left[   \hat{A}^{(2)} [
                                         {\bf r_{1}}, {\bf r_{2}};E] -
                              \left(     \lim_{r_{1}^{\prime} \rightarrow 0}
                                   \frac{\hat{A}^{(2)} [{\bf r_{1}^{\prime}},
                                                 {\bf r_{2}};E]}
                                   { \ln r_{1}^{\prime} }
                              \right)
                         \ln r_{1}
                 \right]  \nn \\
               \lefteqn{    \lim_{r_{1} \rightarrow 0}
              r_{1} \hat{A}^{(3)} [{\bf r_{1}}, {\bf r_{2}};E]  } \nn \\
       & &      =   - \frac{\lambda}{2 \pi}
                     \lim_{r_{1} \rightarrow 0}
                       \left[
                       \hat{A}^{(3)} [{\bf r_{1}}, {\bf r_{2}};E] +
                        r_{1} \frac{\partial}{\partial r_{1}}
                        \hat{A}^{(3)} [{\bf r_{1}}, {\bf r_{2}};E]
                        \right]
\eeqs
where $\lambda$ is a self-adjoint extension parameter,
bound state energies are straightforwardly obtained
\beqs
       B^{(2)} = \frac{2}{M}
                 e^{ 2 ( \frac{\pi}{\lambda} - \gamma)}, \nn \\
       B^{(3)} = \frac{2 \pi^{2}}{M \lambda^{2}}
\eeqs
where $\gamma$ is a Euler constant.
By using (21) the denominators of the right-hand side of
Eq.(18) become
\beqs
      \frac{1}{v} + \frac{M}{\pi} K_{0}[\sqrt{2ME} \epsilon]
      = - \frac{M}{2 \pi} (\ln E - \ln B^{(2)})
        + \frac{1}{v}
        - \frac{M}{\pi} \ln \epsilon
        - \frac{M}{\lambda},   \nn \\
      \frac{1}{v} + \frac{M}{2 \pi \epsilon} e^{
      -\sqrt{2ME} \epsilon}
      = - \frac{M}{\pi} \left(
                        \sqrt{\frac{ME}{2}} -
                        \sqrt{\frac{MB^{(3)}}{2}}
                        \right)
         + \frac{1}{v} + \frac{M}{2 \pi \epsilon}
         - \frac{M}{\lambda}.
\eeqs
Since $B^{(d)}$ must be a pole of $\hat{A}^{(d)}[{\bf
r_{1}}, {\bf r_{2}};E]$ the relations
between the self-adjoint extension parameter
$\lambda$ and the bare coupling constant $v$ are easily derived:
\beqs
     \frac{1}{\lambda} = \frac{1}{Mv}
                   - \frac{1}{\pi} \ln \epsilon
     \;\;\;\;\;\;\;\mbox{for $d = 2$}    \nn \\
     \frac{1}{\lambda} = \frac{1}{Mv} +
                   \frac{1}{2 \pi \epsilon}
     \;\;\;\;\;\;\;\mbox{for $d = 3$}.
\eeqs
The relations of $\lambda$ with $v$ which were derived
in Ref.[4] are
\beqs
      \frac{1}{\lambda} = \frac{1}{Mv}
                      + \frac{1}{\pi} \ln \frac{\Lambda}{2}
                      + \frac{\gamma}{\pi}
                       \;\;\;\;\;\;\;\mbox{for $d = 2$}   \nn \\
      \frac{1}{\lambda} = \frac{1}{Mv}
                       + \frac{\Lambda}{\pi^{2}}
                       \;\;\;\;\;\;\;\mbox{for $d = 3$}
\eeqs
where $\Lambda$ is cut-off parameter of a certain divergent
integral. So Eq.(23) coincides
with Eq.(24) if the relations
\beqs
       \epsilon = \frac{2}{\Lambda} e^{-\gamma}
       \;\;\;\;\;\;\;\mbox{for $d = 2$}  \nn \\
       \epsilon = \frac{\pi}{2 \Lambda}
       \;\;\;\;\;\;\;\mbox{for $d = 3$}
\eeqs
are satisfied. By combining Eqs.(18), (22) and (23) the
energy-dependent Green functions are obtained
\beqs
       \hat{A}^{(2)}[{\bf r_{1}}, {\bf r_{2}};E]
       = \frac{2M/\pi}{\ln E - \ln B^{(2)}}
         K_{0}[\sqrt{2ME} r_{1}] K_{0}[\sqrt{2ME}
         r_{2}],  \nn \\
       \hat{A}^{(3)}[{\bf r_{1}}, {\bf r_{2}};E]
       = \frac{(M/2 \pi)}{\sqrt{2ME} - \sqrt{2MB^{(3)}}}
         \frac{e^{-\sqrt{2ME}(r_{1} + r_{2})}}{r_{1}r_{2}}.
\eeqs
{}From Eq.(26) and the fact that a bound state is a residue
of a energy-dependent Green function, the bound states are
\beqs
        \phi_{B}^{(2)} (r) = \sqrt{\frac{2MB^{(2)}}{\pi}}
                     K_{0}[\sqrt{2MB^{(2)}} r],  \nn \\
        \phi_{B}^{(3)} (r) = \left( \frac{MB^{(3)}}{2 \pi^2}
                         \right)^{\frac{1}{4}}
                         \frac{e^{-\sqrt{2MB^{(3)}}r}}
                              {r}.
\eeqs
Although the time-dependent propagator $G^{(2)}[{\bf
r_{1}}, {\bf r_{2}};t]$ which is a inverse Laplace
transform of $\hat{G}^{(2)}[{\bf r_{1}}, {\bf r_{2}};E]
$ is not well-defined because of $\ln E$ in the
denominator of Eq.(26),  by using the formula\cite{O}
\beq
    L^{-1} \left(\frac{e^{-k\sqrt{E}}}
                      {\sqrt{E} + a } \right)
    = \frac{1}{\sqrt{\pi t}} \exp \left(
                                   -\frac{k^{2}}
                                      {4t}
                                  \right)
       - a e^{ak + a^{2} t}
         erfc(a \sqrt{t} + \frac{k}{2\sqrt{t}})
\eeq
where $L^{-1}$ is a inverse Laplace transform
operator and $erfc(z)$ is usual error function
\beq
     erfc(z) =
     \frac{2}{\sqrt{\pi}} \int_{z}^{\infty}
     e^{-t^{2}} dt,
\eeq
$G^{(3)} [{\bf r_1}, {\bf r_2};t]$ is easily calculated:
\beqs
\lefteqn{ G^{(3)}[{\bf r_{1}}, {\bf r_{2}};t] = } \nn \\
 & & \lefteqn{     G_{0}^{(3)}[{\bf r_{1}}, {\bf r_{2}};t]
       + \left(
               \frac{M}{8 \pi^{3} t}
          \right)^{\frac{1}{2}}
          \frac{1}{r_{1} r_{2}}
           e^{- \frac{M}{2t}
               (r_{1} + r_{2})^{2}}}  \nn \\
     & & \lefteqn{    + \left(
                \frac{MB^{(3)}}{2 \pi^{3}}
           \right)^{\frac{1}{2}}
           \frac{1}{r_{1} r_{2}}
           e^{-\frac{M}{2t}
                (r_{1} + r_{2})^{2}} } \nn \\
       & & \times   \int_{0}^{\infty}du
           \exp \left(
                      -u^{2} +
                             (
                               2\sqrt{B^{(3)}t}
                             - \sqrt{\frac{2M}{t}}
                              (r_{1} + r_{2})
                              ) u
                 \right).
\eeqs
If one uses the formula
\beq
     \frac{d^{n+1}}{dz^{n+1}}
      erf(z) = (-1)^{n} \frac{2}
                           {\sqrt{\pi}}
               H_{n}(z) e^{-z^{2}}
\eeq
where $erf(z) \equiv 1 - erfc(z)$ and $H_{n}$ is usual
Hermite polynomial, it is directly proved that
$G^{(3)}[{\bf r_{1}}, {\bf r_{2}};t]$ obeys the boundary
condition
\beq
    \lim_{r_{1} \rightarrow 0}
    r_{1} G[{\bf r_{1}}, {\bf r_{2}};t]
   = -\frac{\lambda}{2 \pi}
     \lim_{r_{1} \rightarrow 0}
         \left(
               1 + r_{1} \frac{\partial}
                           {\partial r_{1}}
          \right)
          G[{\bf r_{1}}, {\bf r_{2}};t].
\eeq
Of course the Feynman propagator $K^{(3)}[{\bf r_{1}},
{\bf r_{2}};t]$ is obtained by inserting Eq.(30) to
Eq.(4).

\section{two-dimensional spin-1/2 AB problem}

In this section the spin-1/2 AB problem will be
formulated by the path integral method.
The Hamiltonian for the spin-0 AB problem is
\beq
    H_{B} = \frac{1}{2M} ({\bf p} - e{\bf A})^{2}
            + \frac{M \omega^{2}}{2} {\bf r}^{2}
\eeq
where
\beq
     eA_{i} = \alpha \varepsilon_{ij} \frac
              {r_{j}}{r^{2}}
\eeq
and $\alpha$ is a magnetic flux. The harmonic oscillator
potential is included in the Hamiltonian (33) to make the
energy spectrum discrete. The time-dependent Brownian
propagator for the Hamiltonian (33) is well-known\cite{P, Q}
\beq
     G_{B}[{\bf r_{1}}, {\bf r_{2}};t] =
     \sum_{m = -\infty}^{\infty} G_{B,m}^{\omega}
                [r_{1}, r_{2};t] e^{im(\theta_{1}
                                     -  \theta_{2})}
\eeq
where
\beq
      G_{B,m}^{\omega} [r_{1}, r_{2};t] =
       \frac{M \omega}{2 \pi \sinh \omega t}
      e^{-\frac{M \omega}{2} \coth \omega t
      (r_{1}^{2} + r_{2}^{2})}
      I_{\mid m + \alpha \mid}
       \left(
             \frac{M \omega}{\sinh \omega t}
             r_{1} r_{2}
       \right)
\eeq
and $I_{\nu}$ is usual modified Bessel function.

In order to discuss the spin-1/2 AB problem let us
start with the Galilean limit of the Dirac equation as
given in Ref.[1]
\beq
   \left(
         \begin{array}{cc}
         E & -(\Pi_{1} - is\Pi_{2}) \\
        -(\Pi_{1} + is \Pi_{2}) & 2M
         \end{array}
    \right)
    \left(
          \begin{array}{cc}
           \psi_{1}  \\
           \psi_{2}
           \end{array}
     \right)
             = 0
\eeq
where
\beqs
     \Pi_{i} = -i \partial_{i} - e A_{i} \nn \\
     s = \left\{ \begin{array}{11}
                  1    & \mbox{for spin   up}\\
                 -1    & \mbox{for spin   down}
                  \end{array}
         \right..
\eeqs
After including the harmonic oscillator potential
by letting $E \rightarrow E - \frac{1}{2} M \omega^{2}
r^{2}$ the Hamiltonian for $ \psi_{1}$ becomes
\beq
        H_{F} = H_{B} + v \delta({\bf r})
\eeq
where
\beq
        v = \frac{\alpha s}{2M}
            \lim_{r \rightarrow 0}
            \frac{1}{r}.
\eeq
Thus the spin effect in AB problem is to include the
delta-function potential in Hamiltonian which is
interpreted as a Zeeman interaction of the spin with
the magnetic field.

By taking a Laplace transformation to Eq.(36) one can
obtain the energy-dependent Green function for spin-0
AB problem
\beq
    \hat{G}_{B}^{\omega} [{\bf r_{1}}, {\bf r_{2}};E]
    = \sum_{m} \hat{G}_{B,m}^{\omega}[r_{1},
                           r_{2};E]
       e^{im(\theta_{1} - \theta_{2})}
\eeq
where
\beqs
\lefteqn{  \hat{G}_{B,m}^{\omega}[r_{1}, r_{2};E]
     =  \frac{1}{2 \pi \omega r_{1} r_{2}}
       \frac{\Gamma \left(
                          \frac{1 + \mid m+\alpha \mid
                                 + \frac{E}{\omega}}
                                {2}
                    \right) }
             {\Gamma (1 + \mid m+\alpha\mid)} }  \nn \\
  & & \times         W_{-\frac{E}{2\omega},
          \frac{\mid m + \alpha \mid}{2}}
         [M \omega \; max(r_{1}^{2}, r_{2}^{2})]
          M_{-\frac{E}{2\omega},
          \frac{\mid m + \alpha \mid}{2}}
          [M \omega \; min(r_{1}^{2}, r_{2}^{2})]
\eeqs
and $W_{a, b}[z]$ and $M_{a, b}[z]$ are usual Whittaker's
functions. As stated before in Sec.1 $\hat{G}_{B,m}^{\omega}
[r_{1}, 0;E]$ is not well-defined. For the later use
let us calculate the $\hat{G}_{B,m}^{\omega = 0}[r_1, r_2;E]$.
By using the various
asymptotic properties of the Whittaker's functions and
the asymptotic formula
\beq
    \lim_{\omega \rightarrow 0}
     \frac{\Gamma \left(
                        \frac{1 + \frac{E}{\omega}}{2}
                       +\frac{\mid m+ \alpha \mid}{2}
                  \right) }
           {\Gamma \left(
                        \frac{1 + \frac{E}{\omega}}{2}
                       -\frac{\mid m+ \alpha \mid}{2}
                   \right)}
      \approx
              \frac{1}{2^{\mid m+ \alpha \mid}}
                     \left(
                           \frac{E}{\omega}
                     \right)^{\mid m+ \alpha \mid}
\eeq
$\hat{G}_{B,m}^{\omega = 0}[r_{1}, r_{2};E]$ is
\beq
     \hat{G}_{B,m}^{\omega=0}[r_{1}, r_{2};E]
    = \frac{M}{\pi}
      K_{\mid m+ \alpha \mid}(\sqrt{2ME} max(r_{1}, r_{2}))
      I_{\mid m+ \alpha \mid}(\sqrt{2ME} min(r_{1}, r_{2})).
\eeq
By applying the Eq.(14) to this system one gets
\beq
    \hat{A}_{m}^{\omega}[r_{1}, r_{2};E] \\
     = -\frac{1}{\frac{1}{v} + \hat{G}_{B,m}^{\omega}
                                [\epsilon^{+}, \epsilon;E]}
        \hat{G}_{B,m}^{\omega}[r_{1}, \epsilon;E]
        \hat{G}_{B,m}^{\omega}[\epsilon, r_{2};E]
\eeq
where
\beq
    \hat{A}_{m}^{\omega}[r_{1}, r_{2};E]
    = \hat{G}_{F,m}^{\omega}[r_{1}, r_{2};E]
     -\hat{G}_{B,m}^{\omega}[r_{1}, r_{2};E]
\eeq
and $\hat{G}_{F,m}^{\omega}[r_{1}, r_{2};E]$ is $ m^{th}$
component of energy-dependent Green function of spin-1/2
AB problem in the expansion in terms of the angle variable
\beq
    \hat{G}_{F}^{\omega}[{\bf r_1}, {\bf r_2};E] = \sum_{m}
                           \hat{G}_{F,m}^{\omega}
                            [r_{1}, r_{2};E]
                            e^{im(\theta_{1}
                                  - \theta_{2})}.
\eeq
Because of the nomalizibility condition the application
of self-adjoint extension method must be restricted at
$\mid m+ \alpha \mid < 1$ region.\cite{A, B} So we restrict
ourselves in this region. By inserting Eq.(36) to Eq.(45)
$\hat{A}_m^{\omega} [r_1, r_2;E]$ becomes
\beq
    \hat{A}_{m}^{\omega}[r_{1}, r_{2};E] \\
    = f_{m}^{\omega}[\epsilon, E]
      \frac{W_{-\frac{E}{2\omega},
               \frac{\mid m+ \alpha \mid}{2}}
               (M \omega r_{1}^{2})}
           {r_{1}}
      \frac{W_{-\frac{E}{2\omega},
               \frac{\mid m+ \alpha \mid}{2}}
               (M \omega r_{2}^{2})}
            {r_{2}}
\eeq
where
\beq
      f_{m}^{\omega} (\epsilon, E) \\
    = -\frac{1}{\frac{1}{v} +
               \hat{G}_{B,m}^{\omega}[\epsilon^{+},
                \epsilon;E] }
    \left[
          \frac{1}{2 \pi \omega \epsilon}
          \frac{\Gamma \left(
                            \frac{1 + \mid m+ \alpha \mid
                                   + \frac{E}{\omega}}{2}
                       \right) }
               {\Gamma (1 + \mid m+ \alpha \mid) }
            M_{-\frac{E}{2\omega},
                \frac{\mid m+ \alpha \mid}{2}}
              (M \omega \epsilon^{2})
     \right]^{2}.
\eeq
In order to apply the boundary condition at the origin it
is more convenient to express $\hat{A}_{m}^{\omega}[r_{1},
r_{2};E]$ in terms of the confluent hypergeometric
functions:
\beqs
 \lefteqn{ \hat{A}_{m}^{\omega}[r_{1}, r_{2};E] =
    \frac{f_{m}^{\omega}(\epsilon, E)}
           {\mid m+ \alpha \mid^{2}}
      (M \omega)^{1 + \mid m+ \alpha \mid}
      (r_{1} r_{2})^{\mid m+ \alpha \mid} }  \nn \\
 & & \times     e^{-\frac{M \omega}{2} (r_{1}^{2} + r_{2}^{2})}
     g_{m}^{\omega}(r_{1};E)
                   g_{m}^{\omega}(r_{2};E)
\eeqs
where
\beqs
 \lefteqn{   g_{m}^{\omega}(r;E) = }             \nn \\
   & & \lefteqn{  \frac{1}{(M \omega r^{2})^{\mid m+ \alpha \mid}}
    \frac{\Gamma (1 + \mid m+ \alpha \mid)}
         {\Gamma \left(
                       \frac{1 + \mid m+ \alpha \mid
                              + \frac{E}{\omega} }
                            {2}
                  \right) }
    F(\frac{1 - \mid m+ \alpha \mid + \frac{E}{\omega}}{2}
      \mid 1 - \mid m+ \alpha \mid \mid M\omega r^{2} )}   \nn \\
   & &   - \frac{\Gamma (1 - \mid m+ \alpha \mid)}
           {\Gamma \left(
                         \frac{1 - \mid m+ \alpha + \mid
                               \frac{E}{\omega}}
                              {2}
                   \right) }
     F(\frac{1 + \mid m+ \alpha \mid + \frac{E}{\omega}}{2}
       \mid 1 + \mid m+ \alpha \mid \mid M \omega r^{2} ).
\eeqs
By applying the boundary condition\cite{R} which aries by the
self-adjoint extension of the Hamiltonian
\beqs
 \lefteqn{  \lim_{r_{1} \rightarrow 0} r_{1}^{\mid m+ \alpha \mid}
     \hat{A}_{m}^{\omega}[r_{1}, r_{2};E] =
      \lambda_{m} \lim_{r_{1} \rightarrow 0}
       \frac{1}{r_{1}^{\mid m+ \alpha \mid}} }  \nn \\
& &  \times     \left[
             \hat{A}_{m}^{\omega}[r_{1}, r_{2};E]
             - \left(
                     \lim_{r_{1}^{\prime} \rightarrow 0}
                      r_{1}^{\prime \mid m+ \alpha \mid}
                      \hat{A}_{m}^{\omega}[r_{1}^{\prime},
                                           r_{2};E]
               \right)
               \frac{1}{r_{1}^{\mid m+ \alpha \mid}}
        \right]
\eeqs
where $\lambda_{m}$ is a self-adjoint extension parameter, it
is found that the bound state energy is implicitly determined
by the equation
\beq
    \frac{\Gamma \left(
                       \frac{1 + \mid m+ \alpha \mid
                              + \frac{E}{\omega}}
                             {2}
                 \right) }
          {\Gamma \left(
                        \frac{1 - \mid m+ \alpha \mid
                              - \frac{E}{\omega}}
                              {2}
                  \right) }
     = - \frac{1}{\lambda_{m} (M\omega)^{\mid m+
                                   \alpha \mid}}
         \frac{\Gamma ( 1 + \mid m+ \alpha \mid )}
              {\Gamma ( 1 - \mid m+ \alpha \mid )}.
\eeq
Although Eq.(53) is too complicated to evaluate the
bound state energy explicitly, its limiting feature
is interesting. First in the $\lambda_{m} \rightarrow
0$ or $ \infty $ limit the bound state energies are
explicitly determined as a pole of the gamma functions
\beqs
     \lambda_{m} = 0  \;\;\;\;\; B_{n,m} = (1 + 2n + \mid m+
      \alpha \mid) \omega \nn \\
                \qquad                  n = 0, 1, 2, ... \nn \\
     \lambda_{m} = \infty \;\;\;\;\; B_{n,m} = (1 + 2n - \mid m+
      \alpha \mid) \omega.
\eeqs
These bound state energies coincide with those of
regular and singular solutions given in Eq.(1) of Ref.\cite{S}.
Another interesting case is the case of vanishing harmonic
oscillator potential. This is achieved by using Eq.(43):
\beq
    \lim_{\omega \rightarrow 0} B =
    \frac{2}{M} \left(
                      -\frac{1}{\lambda_{m}}
                      \frac{\Gamma(1 + \mid m+ \alpha \mid)}
                           {\Gamma(1 - \mid m+ \alpha \mid)}
                 \right)^{\frac{1}{\mid m+ \alpha \mid}}.
\eeq
This result also coincides with Eq.(3.13) of Ref.[18].

Now let us calculate the energy-dependent Green function
explicitly for spin-1/2 AB problem. For simplicity we only
consider the $\omega = 0$ case. By using Eq.(43) and the
asymptotic formulae of confluent hypergeometric functions
\beqs
     \lim_{a \rightarrow \infty}
      \frac{F(a\mid b\mid \frac{z}{a})}{\Gamma(b)}
     = z^{\frac{1-b}{2}} I_{b-1}(2 \sqrt{z}) \nn \\
      \lim_{a \rightarrow \infty}
      \Gamma(1+a-b) U(a\mid b\mid \frac{z}{a})
     = 2 z^{\frac{1-b}{2}} K_{b-1}(2 \sqrt{z})
\eeqs
where
\beq
     U(a\mid b\mid z) =
     \frac{\pi}{\sin (\pi b)}
       \left(
            \frac{F(a\mid b\mid z)}
                 {\Gamma(1+a-b) \Gamma(b)}
            - z^{1-b}
              \frac{F(1+a-b\mid 2-b\mid z)}
                   {\Gamma(a) \Gamma(2-b)}
        \right)
\eeq
$\hat{A}_{m}^{\omega=0}[r_{1}, r_{2};E]$ becomes
\beq
     \hat{A}_{m}^{\omega=0}[r_{1}, r_{2};E] =
      C_{m}(\epsilon, E) K_{\mid m+ \alpha \mid}
                           (\sqrt{2ME}r_{1})
                         K_{\mid m+ \alpha \mid}
                           (\sqrt{2ME} r_{2})
\eeq
where
\beq
     C_{m}(\epsilon,E) =
     - \left( \frac{M}{\pi} \right)^{2}
     \frac{I_{\mid m+ \alpha \mid}^{2}(\sqrt{2ME} \epsilon)}
          {\frac{1}{v} + \frac{M}{\pi}
                        K_{\mid m+ \alpha \mid}(\sqrt{2ME}
                                                \epsilon)
                        I_{\mid m+ \alpha \mid} (\sqrt{2ME}
                                                 \epsilon)}.
\eeq
By using the behaviors of the modified Bessel functions for
the small argument
\beqs
    \lim_{z \rightarrow 0} I_{\mid m+ \alpha \mid}(z)
    \sim \frac{ \left( \frac{z}{2} \right)^{\mid m+ \alpha
                                            \mid} }
              {\Gamma(1 + \mid m+ \alpha \mid)}
         \left[
               1 + \frac{ \left(\frac{z}{2} \right)^{2} }
                        {1+ \mid m+ \alpha \mid}
        \right] \nn \\
      \lim_{z \rightarrow 0} K_{\mid m+ \alpha \mid}(z)
      \sim \frac{1}{2}
           \left[
                 \Gamma(\mid m+ \alpha \mid)
                 \left( \frac{z}{2} \right)^{-\mid m+ \alpha
                                             \mid}
                 - \frac{ \left(\frac{z}{2}\right)^{\mid m+
                                           \alpha \mid}}
                        {\Gamma(1 + \mid m+ \alpha \mid)}
            \right]
\eeqs
and Eq.(55) the denominator of the right-hand side of Eq.(
59) is
\beqs
 \lefteqn{ \frac{1}{v} + \frac{M}{\pi}
                 K_{\mid m+ \alpha \mid}(\sqrt{2ME} \epsilon)
                 I_{\mid m+ \alpha \mid}(\sqrt{2ME} \epsilon)  } \nn \\
    & & \lefteqn{ = - \frac{M}{2 \pi \mid m+ \alpha \mid}
        \epsilon^{2 \mid m+ \alpha \mid}
        \frac{\Gamma(1 - \mid m+ \alpha \mid)}
             {\Gamma(1 + \mid m+ \alpha \mid)} }    \nn \\
    & &  \times  \left[  \left(\frac{ME}{2}\right)^{\mid m+ \alpha \mid}
                - \left(\frac{MB}{2}\right)^{\mid m+ \alpha \mid}
                \right]
\eeqs
on the condition that
\beq
    \frac{1}{\lambda_{m}} = -
    \frac{1}{\epsilon^{2 \mid m+ \alpha \mid}}
     \left(
           1 + \frac{2 \pi \mid m+ \alpha \mid}{Mv}
     \right).
\eeq
Eq.(62) is the relation of the self-adjoint extension
parameter with the bare coupling constant in this model.
By combining Eqs.(58), (59) and (61) $\hat{A}_{m}^{\omega=0}[r_{1}, r_2;E]$
becomes
\beqs
 \lefteqn{  \hat{A}_m^{\omega=0}[r_1,r_2;E] = } \nn \\
    & &  \lefteqn{ \frac{2M}{\pi^2} \sin (\mid m+ \alpha \mid \pi)
       \frac{E^{\mid m+ \alpha \mid}}
            {E^{\mid m+ \alpha \mid} - B^{\mid m+ \alpha
                                         \mid} } } \nn \\
    & &  \times K_{\mid m+ \alpha \mid}(\sqrt{2ME} r_1)
                K_{\mid m+ \alpha \mid}(\sqrt{2ME} r_2).
\eeqs
Bound state is easily obtained from Eq.(63)
\beq
    \phi_B(r) = \sqrt{
                      \frac{2MB \sin(\mid m+ \alpha \mid \pi) }
                           {\pi^2 \mid m+ \alpha \mid} }
                K_{\mid m+ \alpha \mid}(\sqrt{2MB}r).
\eeq
{}From Eq.(63) it is easily found that the inverse Laplace
transformation of $ \hat{A}_m^{\omega=0}[r_1, r_2;E]$ is
not well-defined except B = 0(or
$\lambda_m \rightarrow \infty$) case. This case coincides
with the Hagen's delta function limiting procedure.[1]
To calculate the time-dependent propagator explicitly in
this case let us perform the inverse Laplace transform
to $\hat{A}_m^{\omega=0}[r_1, r_2;E]$. The result is
\beqs
 \lefteqn{ \left[
          G_F[{\bf r_1}, {\bf r_2};t] -
          G_B[{\bf r_1}, {\bf r_2};t]
    \right]_{\lambda_m \rightarrow \infty} = }  \nn \\
    & &  \lefteqn{ \sum_{\mid m+ \alpha \mid < 1}
      \frac{M \sin (\mid m+ \alpha \mid \pi) }
           {\pi^2 t}
      e^{- \frac{M}{2t} (r_1^2 + r_2^2)} } \nn \\
  & &  \times      K_{\mid m+ \alpha \mid}
                \left( \frac{Mr_1r_2}{t} \right)
      e^{im(\theta_1 - \theta_2)}
\eeqs
So the Feynman kernel is obtained by using the Eq.(4):
\beqs
 \lefteqn{  K_{F,\lambda_m \rightarrow \infty}^{\omega=0}
     [{\bf r_1}, {\bf r_2};t] =
    \frac{M}{2\pi it} e^{\frac{iM}{2t} (r_1^2 + r_2^2)} } \nn \\
  & &   \lefteqn{     \Bigg[
              \sum_{\mid m+ \alpha \mid > 1}
               e^{im(\theta_1 - \theta_2)-\frac{i\pi}{2}
                  \mid m+ \alpha \mid}
               J_{\mid m+ \alpha \mid}
                 \left( \frac{M}{t}r_1 r_2 \right) } \nn \\
      & &  \lefteqn{       + \sum_{\mid m+ \alpha \mid < 1}
                 e^{im(\theta_1 - \theta_2)}
                 \bigg\{  e^{-\frac{i\pi}{2} \mid m+ \alpha
                            \mid}
                        J_{\mid m+ \alpha \mid}
                    \left( \frac{M}{t} r_1 r_2 \right) } \nn \\
      & &                   + i \sin (\mid m+ \alpha \mid \pi)
                           e^{\frac{i\pi}{2} \mid m+ \alpha
                              \mid}
                           H_{\mid m+ \alpha \mid}^{(1)}
                             \left( \frac{M}{t} r_1r_2 \right)
                  \bigg\}
                      \Bigg]
\eeqs
where $J_{\nu}$ and $H_{\nu}^{(1)}$ are usual Bessel and Hankel
functions. Finally by using the relation
\beq
    H_{\nu}^{(1)}(z) =
    \frac{i}{\sin \nu \pi}
     \left( e^{-\nu \pi i} J_{\nu}(z) - J_{-\nu}(z) \right)
\eeq
Eq.(66) becomes
\beqs
  \lefteqn{ K_{F,\lambda_m \rightarrow \infty}^{\omega=0}
     [{\bf r_1}, {\bf r_2};t] =
      \frac{1}{2 \pi it}
     e^{\frac{iM}{2t} (r_1^2 + r_2^2)} }  \nn \\
  & &   \lefteqn{     \Bigg[
            \sum_{\mid m+ \alpha \mid > 1}
            e^{im(\theta_1 - \theta_2) - \frac{i\pi}{2}
               \mid m+ \alpha \mid}
            J_{\mid m+ \alpha \mid}
              \left( \frac{M}{t} r_1 r_2 \right) }  \nn \\
  & &              +  \sum_{\mid m+ \alpha \mid < 1}
            e^{im(\theta_1 - \theta_2) + \frac{i\pi}{2}
                \mid m+ \alpha \mid}
            J_{-\mid m+ \alpha \mid}
              \left(  \frac{M}{t} r_1 r_2 \right)
           \Bigg] .
\eeqs
So when $\mid m+ \alpha \mid < 1$ olny the singular solution is
contributed to the Feynman propagator which is the property of
the spin-1/2 AB problem when the delta function potential is
treated as a limit of the infinitesimal radius.

\section{Conclusion}

The delta-function potentials in two- and three- dimensional
quantum mechanics are analyzed by using the path-integral
formalism. The energy-dependent Green functions for free
particle plus delta-function potential systems are expressed
in terms of the self-adjoint extension parameter. It is found
that the time-dependent propagator of the two-dimensional case
is not well-defined while that of three-dimensional case is
well-defined.

By applying this method to the two-dimensional spin-1/2
AB problem the energy-dependent Green function for this
system is explicitly calculated. It is found that the
time-dependent propagator is not well-defined except
only one special value of the self-adjoint extension parameter
$\lambda_m = \infty$. At this value only the singular
solutions are contributed to the time-dependent propagator
which is the
well-known property of the spin-1/2 AB problem when
the delta function potential is treated as a limit of
the infinitesimal radius. The existence of the energy-
dependent and time-dependent propagators in this limit
may shed light on the proof of the conjecture that there is
only one Hamiltonian that is correct limit as
solenoid radius goes to zero.

\bigskip

\noindent {\em ACKNOWLEDGMENTS}

I wish to thank to the
Department of Physics and Astronomy in the University
of Rochester for hospitality. Especially I am very
grateful to C.R.Hagen for introducing spin-1/2 AB problem.
This work is performed in the course of the post-
doctoral program supported from Korean Science and
Engineering Foundation.

\end{document}